\documentclass[a4paper]{article}

\usepackage{graphicx}
\usepackage{natbib}

\newcommand{\ica}{\sc{FastICA}}
\newcommand{\lsim}{\,\lower2truept\hbox{${<\atop\hbox{\raise4truept\hbox{$\sim$}}}$}\,}
\newcommand{\gsim}{\,\lower2truept\hbox{${>\atop\hbox{\raise4truept\hbox{$\sim$}}}$}\,}

\begin{document}

\title{Cosmic Microwave Background Polarisation: foreground contrast 
and component separation}

\author{Carlo Baccigalupi \and
SISSA/ISAS, Via Beirut 4, Trieste, 34014 Italy and \and
LBNL, 1 Cyclotron Road, Berkeley, CA 94720, USA \and
{\tt bacci@sissa.it, bacci@materia.lbl.gov}}

\maketitle

\begin{abstract}
We evaluate the expected level of foreground contamination to the 
cosmic microwave background (CMB) polarised radiation, focusing on 
the diffuse emission from our own Galaxy. In particular, we perform a 
first attempt to simulate an all sky template of polarised emission 
from thermal dust. This study indicates that the 
foreground contamination to CMB $B$ modes is likely to be relevant 
on all frequencies, and even at high Galactic latitudes. \\
We review the recent developments in the design of data analysis 
techniques dedicated to the separation of CMB and foreground 
emissions in multi-frequency observations, exploiting their 
statistical independence. We argue that the high quality and detail of 
the present CMB observations represent an almost ideal statistical 
dataset where these algorithms can operate with excellent performance. 
We explicitly show that the recovery of CMB $B$ modes is possible even 
if they are well below the foreground level, working at the arcmin. 
resolution at an almost null computational cost. This capability well 
represents the great potentiality of these new data analysis techniques, 
which should be seriously taken into account for implementation in present 
and future CMB observations. 
\end{abstract}

\section{Introduction}
\label{intro}

Among all cosmological observables, the Cosmic Microwave Background (CMB) 
polarisation stores cosmological perturbations in the smartest way. 
The Stokes parameters $Q$ and $U$ are non-locally combined to get the $E$ 
and $B$ modes, featuring opposite parity relations; as for total intensity 
($T$), $E$ modes are activated by all cosmological perturbations, and indeed 
the two are strongly correlated ($TE$); on the other hand, 
$B$ modes are excited by vectors and tensors only; since no significant power 
is experted from vectors, a detection of $B$ modes would be a strong indication 
of the presence of cosmological gravitational waves. 
For a comprehensive review of the CMB physics, see Hu et al. (1998) and 
references therein. 

On the observation side, we are right now in the epoch in which the CMB is 
revealing its finest structure. The Wilkinson Microwave Anisotropy Probe (WMAP) 
satellite\footnote{\tt map.gsfc.nasa.gov} released recently the first year 
CMB data, for $T$ and $TE$, unveiling all sky anisotropies with about $20$ 
arcmin. resolution on a frequency range between $22$ and 90 GHz 
\citep[see][and references therein]{BENNE}; balloon-borne 
and ground-based $T$ observations cover angular scales from a few arcminutes 
to tens of degrees, and also aim at measuring CMB $E$ and $B$ modes 
\citep[see][for a first $E$ mode detection]{KOVAC}. The {\sc Planck} 
satellite\footnote{\tt astro/estec.esa.nl/SA-general/Projects/Planck} 
will provide total intensity and polarisation full sky maps with 
$5'$ resolution and a sensitivity of a few $\mu$K, on nine 
frequency channels between 30 and 857 GHz. 

\begin{figure}
\centering
\includegraphics[angle=90,width=6cm,height=5cm]{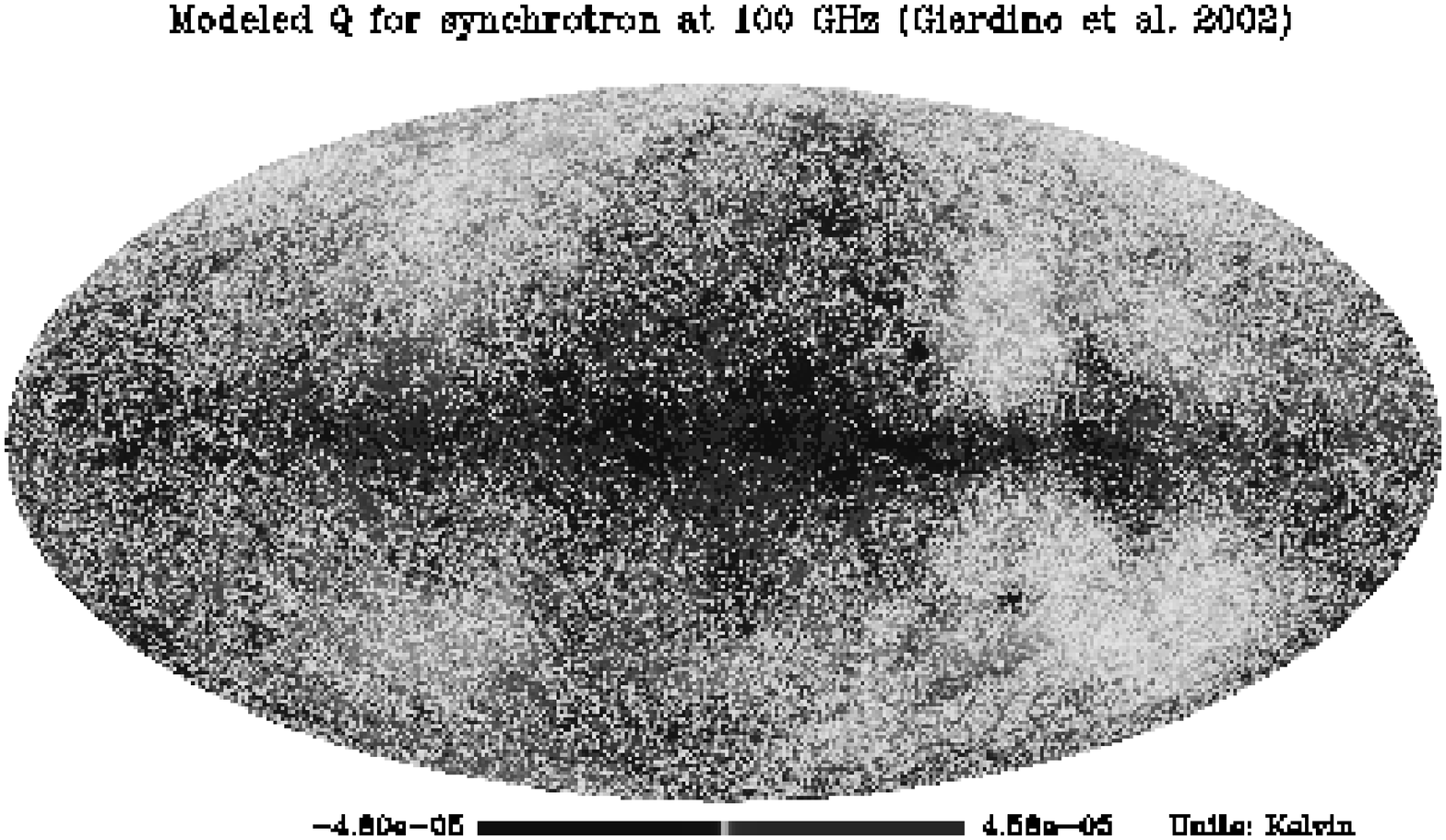}
\includegraphics[angle=90,width=6cm,height=5cm]{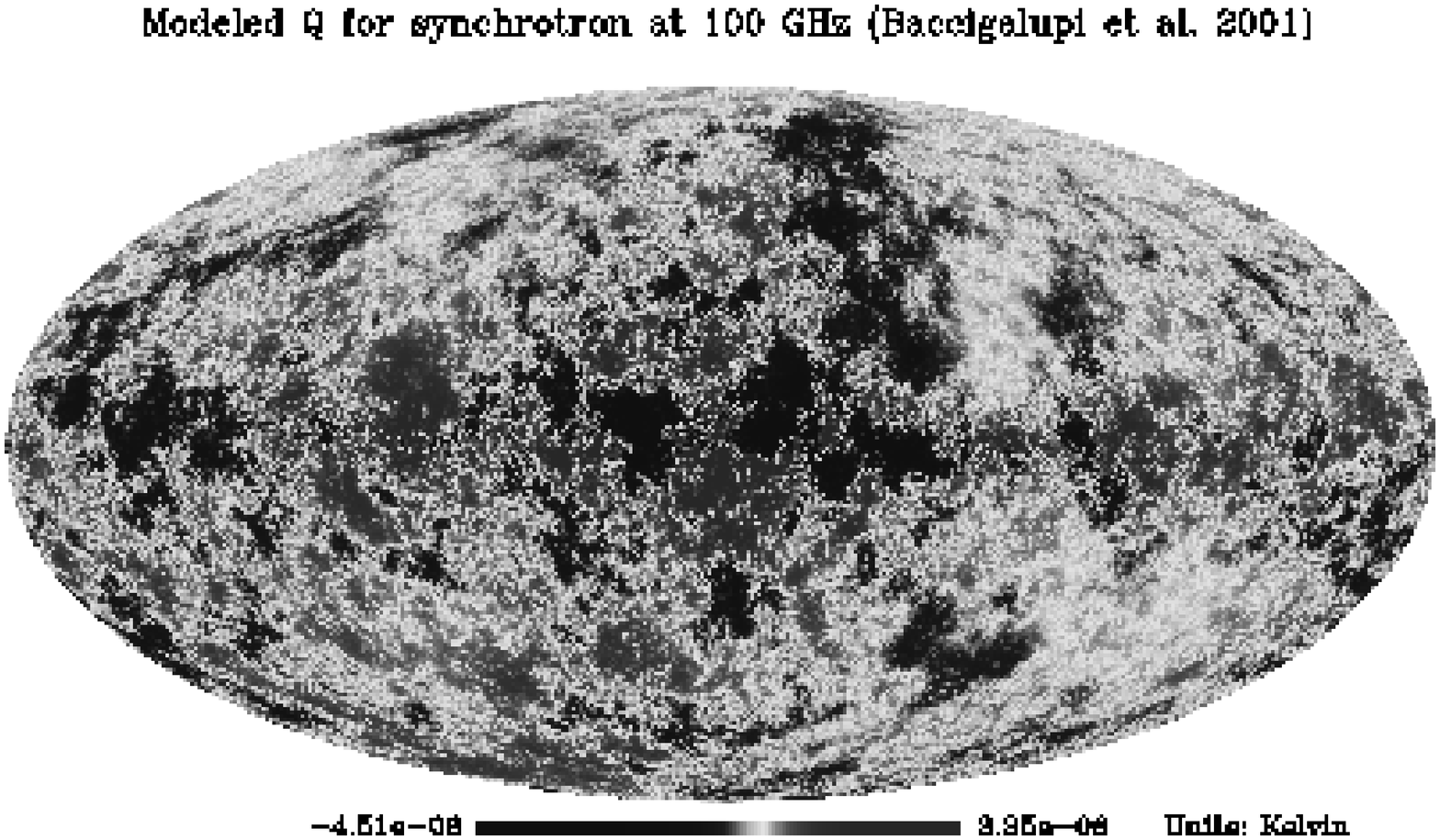}
\includegraphics[angle=90,width=6cm,height=5cm]{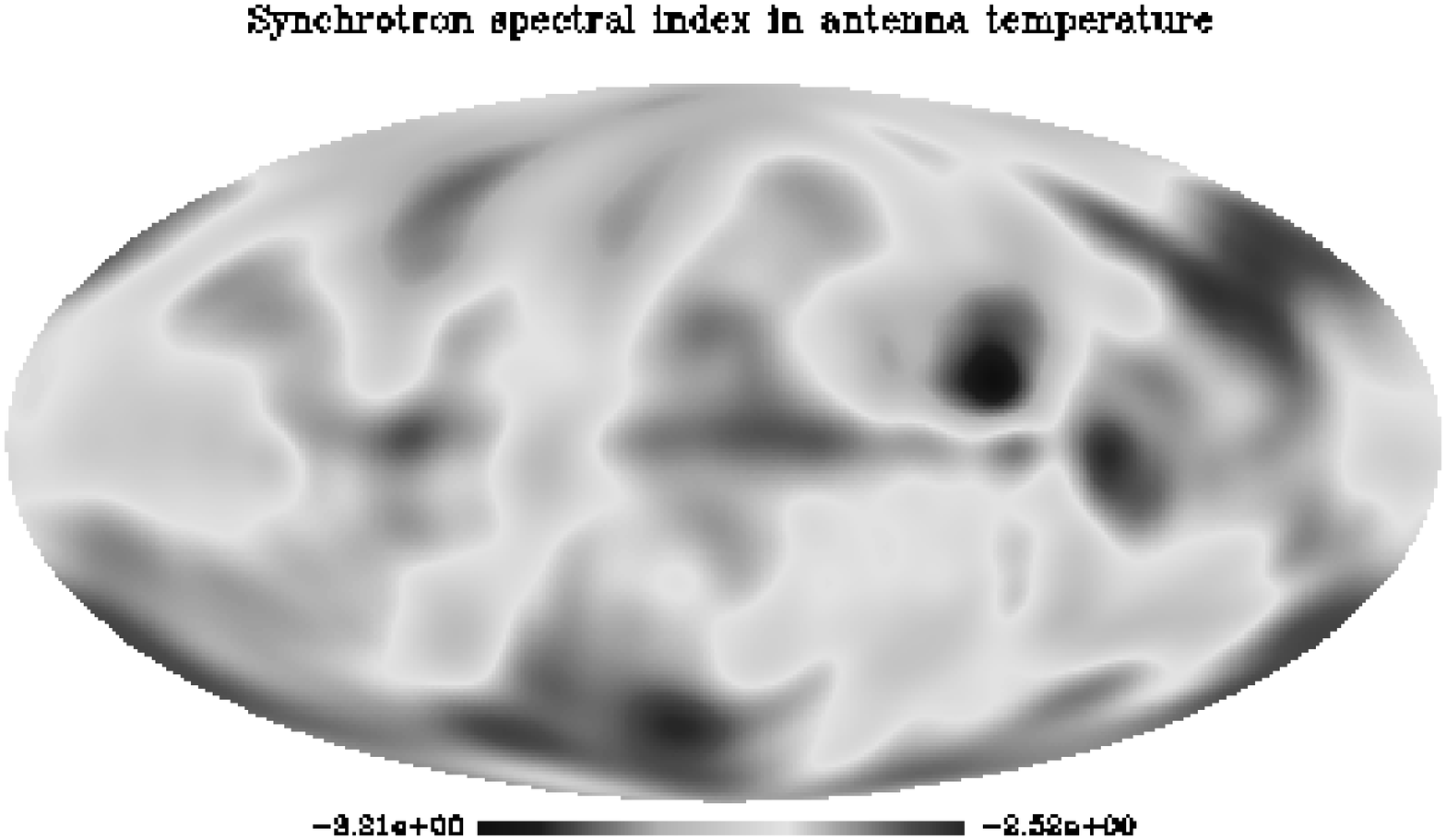}
\includegraphics[angle=90,width=6cm,height=5cm]{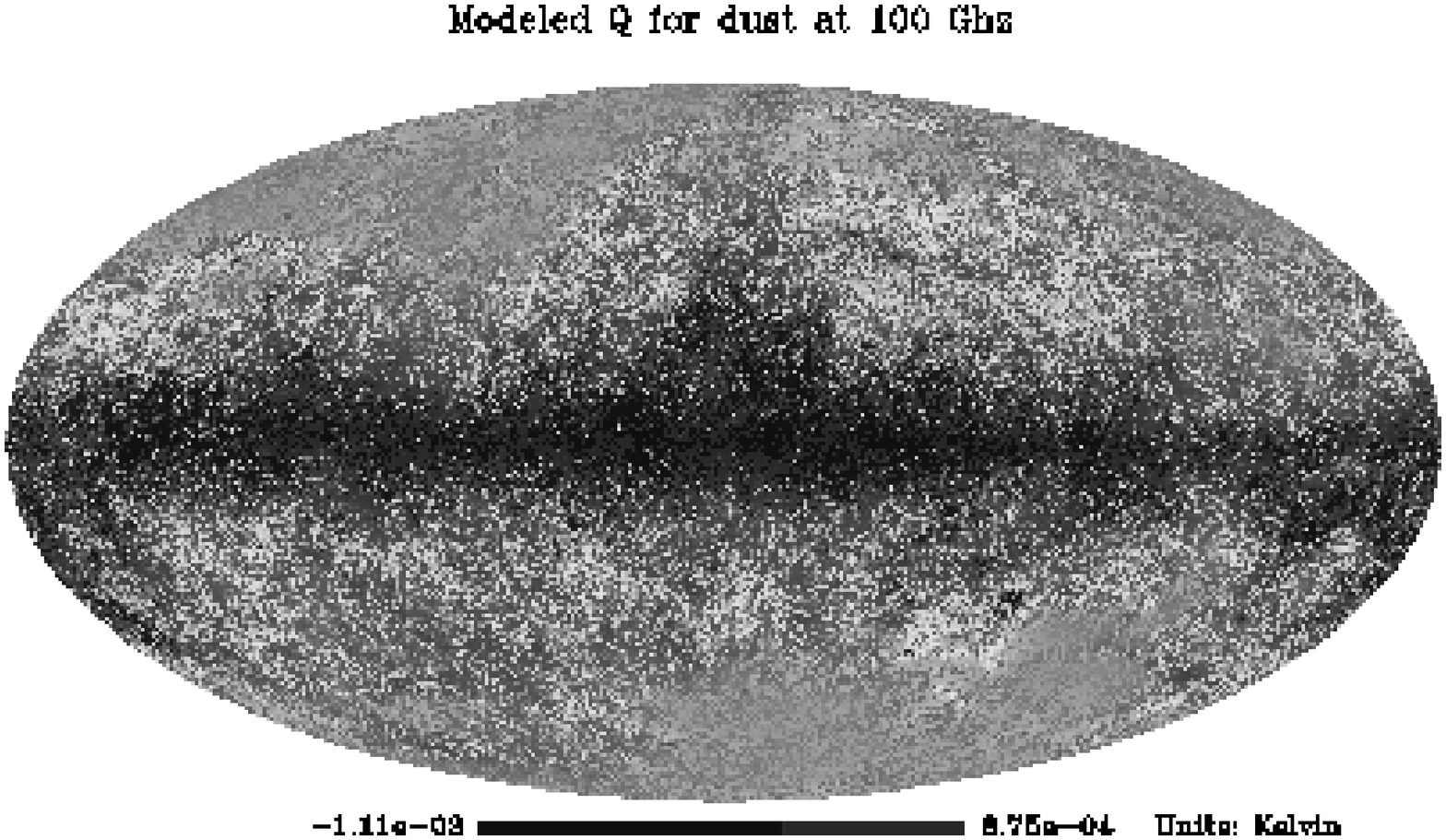}
\caption{Upper panels: simulated templates for the synchrotron $Q$ Stokes 
parameter. Lower left panel: synchrotron spectral index template. 
Lower right panel: $Q$ simulated template for the dust.}
\label{syndustsky}
\end{figure}

This work focuses on the foreground contamination to the CMB polarised 
emission. This issue is gathering greater and greater attention as the 
importance of the CMB polarisation measurements increases 
\cite[see][for reviews]{DEZOT}. In Section \ref{foremis} we estimate the 
expected strength of foreground polarised emission with respect to the CMB; 
in Section \ref{compsep} we describe the recent progresses in the design of 
data analysis tools designed to reduce such a contamination; in Section 
\ref{conclud} we make some concluding remarks. 

\section{Foreground contamination to the CMB}
\label{foremis}

We focus on the diffuse emission from our own Galaxy. Indeed, 
at microwave frequencies, the contamination from Galactic, extra-Galactic 
radio and infrared sources \citep[see][and references therein]{DEZOT}, as 
well as the polarised Sunyaev-Zel Dovich emission \cite[see][proceedings 
of this conference]{WHITE}, are likely to be relevant for the damping tail 
of the CMB power spectrum, i.e. at arcmin. angular scales, corresponding 
to multipoles $\ell$ of the order 1000, which sets the limit of validity 
for our analysis. 

The main diffuse foreground at low microwave frequencies, say $\lsim 100$ 
GHz is expected from Galactic synchrotron. 
The available high resolution data in the radio band, approximately $10$ 
arcmin., are at low and medium Galactic latitudes, $|b|\lsim 20^{\circ}$ 
\citep{DUNCA,UYANI}; degree resolution data \citep{BROUW} cover 
about half of the sky and reach high latitudes. These templates have been 
used to evaluate the angular power spectrum of synchrotron in the radio 
band, up to arcmin. scale 
\citep[][proceedings of this conference]{TUCCI1,BACCI2,GIARD,TUCCI2,DEOLI}. \\
\citet{TUCCI1} estimated $C_{\ell}^{E,B}\propto\ell^{-1.5\div -2}$, 
regardless of the Galactic region considered; note that the 
same behavior was quoted for the first time by \citet{TEGMA}. 
\citet{BACCI2} found the same result, regardless also 
of the Galactic latitude, up to the highest probed $|b|\simeq 20^{\circ}$; 
they also estimated the level of Faraday depolarisation to be important, but 
not such to mask substantially the true synchrotron emission; moreover, 
they estimated the power on super-degree angular scales, corresponding 
to multipoles $\ell\le 200$, using the \citet{BROUW} data, finding a 
steeper behavior, $C_{\ell}\propto\ell^{-3}$. \citet{GIARD} estimated 
the power spectrum of the cosine and sine of the polarisation angle up 
to the arcmin. scale, and out of that constructed a spatial template 
assuming Gaussian distribution; the resulting pattern was then used to 
simulate templates for $Q$ and $U$ out of the total intensity Haslam map 
at 408 MHz, assumed to be synchrotron dominated and theoretically polarised 
at $75\%$; they also estimated the synchrotron spectral index as inferred by 
multi-frequency radio observations. \citet{TUCCI2} extended the power 
spectrum analysis up to a few thousand multipoles, and \citet{DEOLI} performed 
a detailed study of the $E$ and $B$ modes in the large scale data. 

The CMB contamination from diffuse Galactic dust is much more 
uncertain. The main emission mechanism arises from the thermal emission 
of magnetized dust grains, which get locally aligned by the Galactic 
magnetic field \citep{LAZAR}. Moreover, indications of a non-thermal dust 
emission, possibly connected with rapid rotations of the grains and 
extending to much lower frequencies, has been conjectured in several works, 
although in the WMAP data there is no evidence of such component 
\cite[see][and references therein]{BENNE}. In total intensity, a $6'$ 
resolution all sky template of dust emission is available at 
100 $\mu$m, and on the basis of a correct interpretation it can be scaled 
at microwave frequencies and compared with the CMB \citep[see][and 
references therein]{FINKB}. In polarisation, a great achievement 
has been obtained by Archeops 
\citep[see][proceedings of this conference]{PONTH}, which measured a 
$3-5\%$ polarised dust signal on the Galactic plane. On the basis of these 
two ingredients, we propose here a naively simulated template of the 
polarised dust emission, which agrees remarkably well with previous 
estimates \citep{LAZAR,PRUNE}. We assume perfect correlation between 
total and polarised intensity and a $5\%$ polarisation degree. Then we 
assume the same polarisation angle pattern of the synchrotron emission 
\citep{GIARD}. That is certainly quite drastic, but it corresponds to 
the assumption that the Galactic magnetic field is 100\% efficient 
imprinting the polarisation direction to the dust and synchrotron 
emission. 

\begin{figure}
\centering
\includegraphics[width=6cm,height=4.3cm]{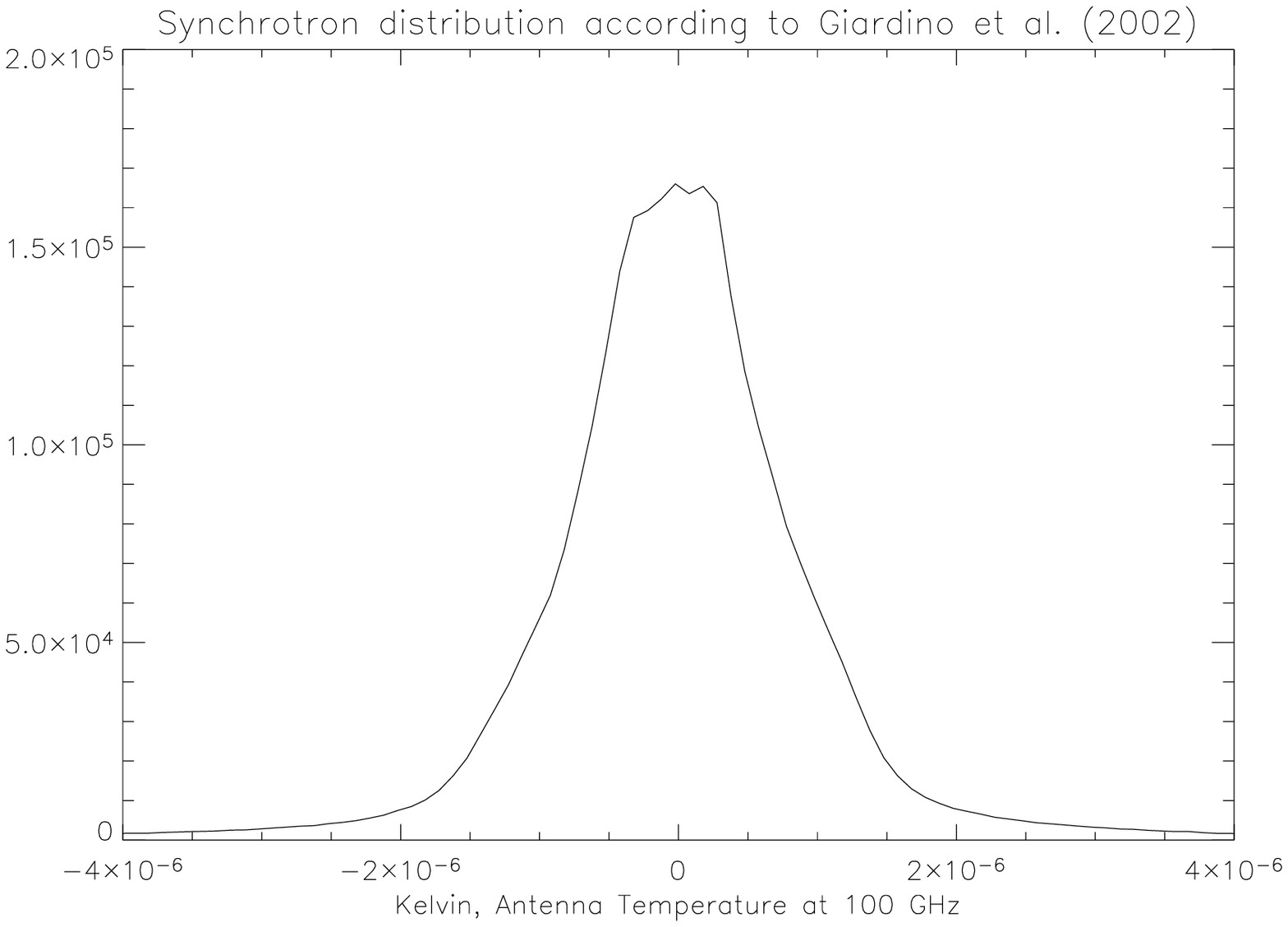}
\includegraphics[width=6cm,height=4.3cm]{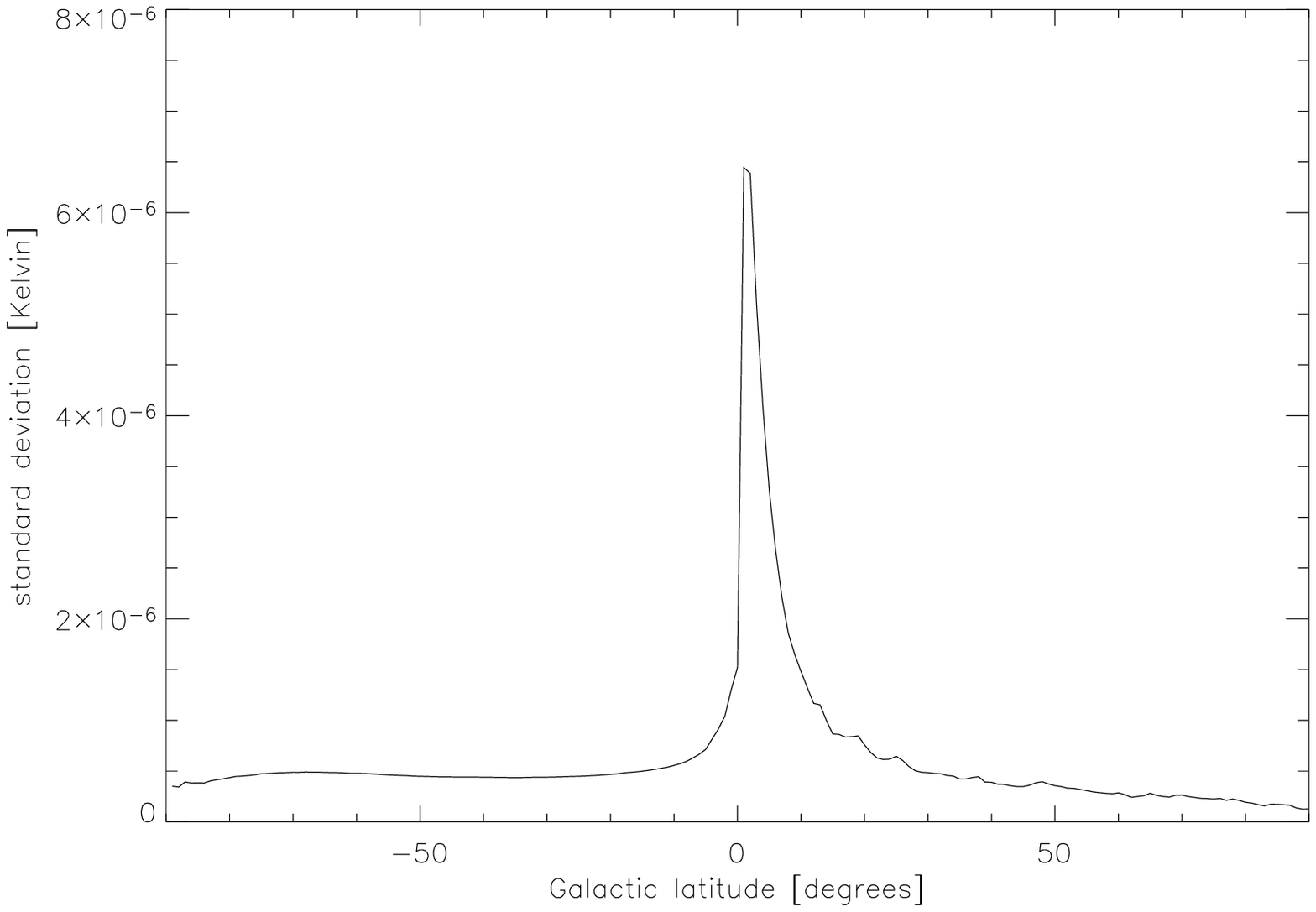}
\includegraphics[width=6cm,height=4.3cm]{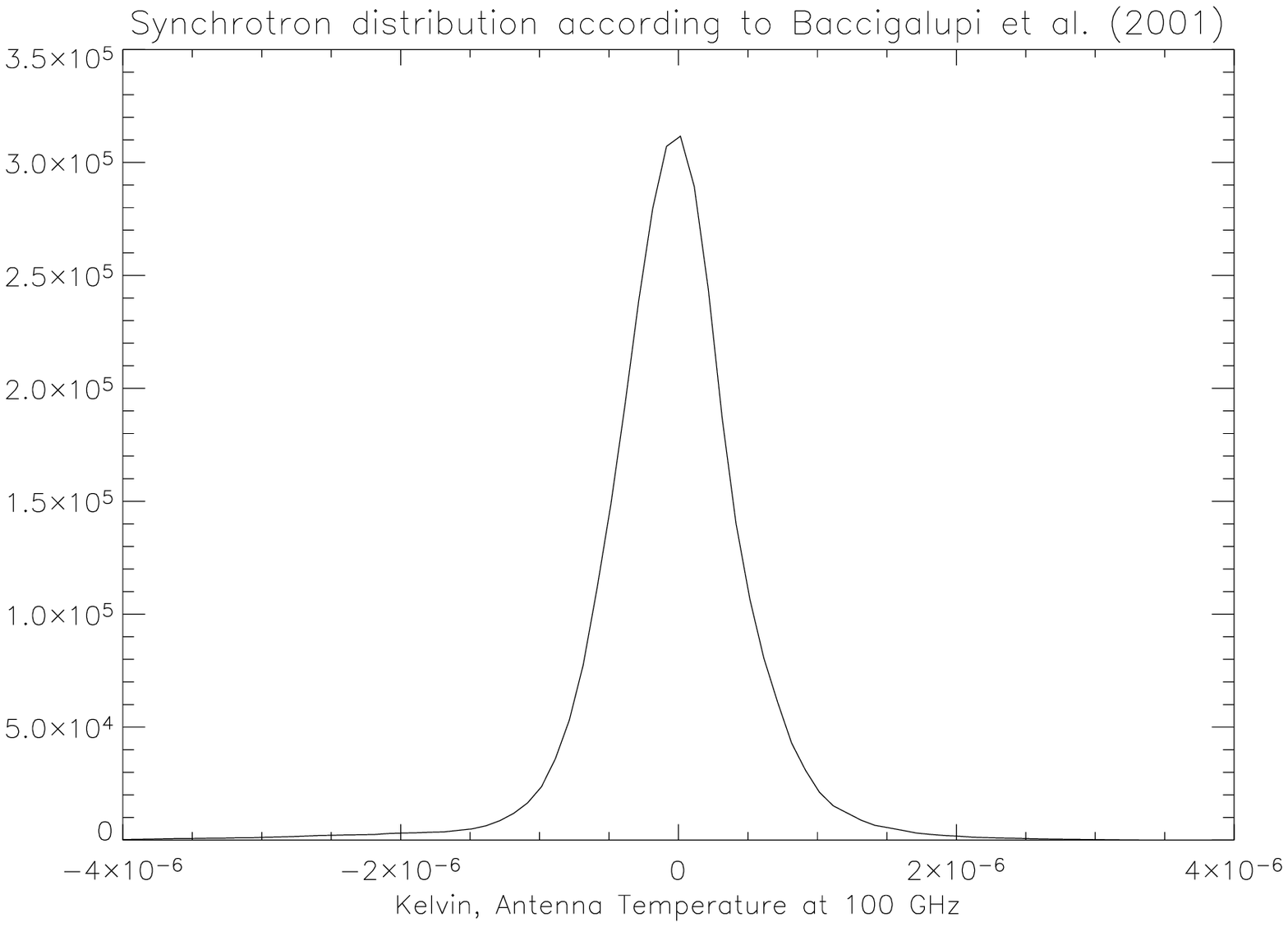}
\includegraphics[width=6cm,height=4.3cm]{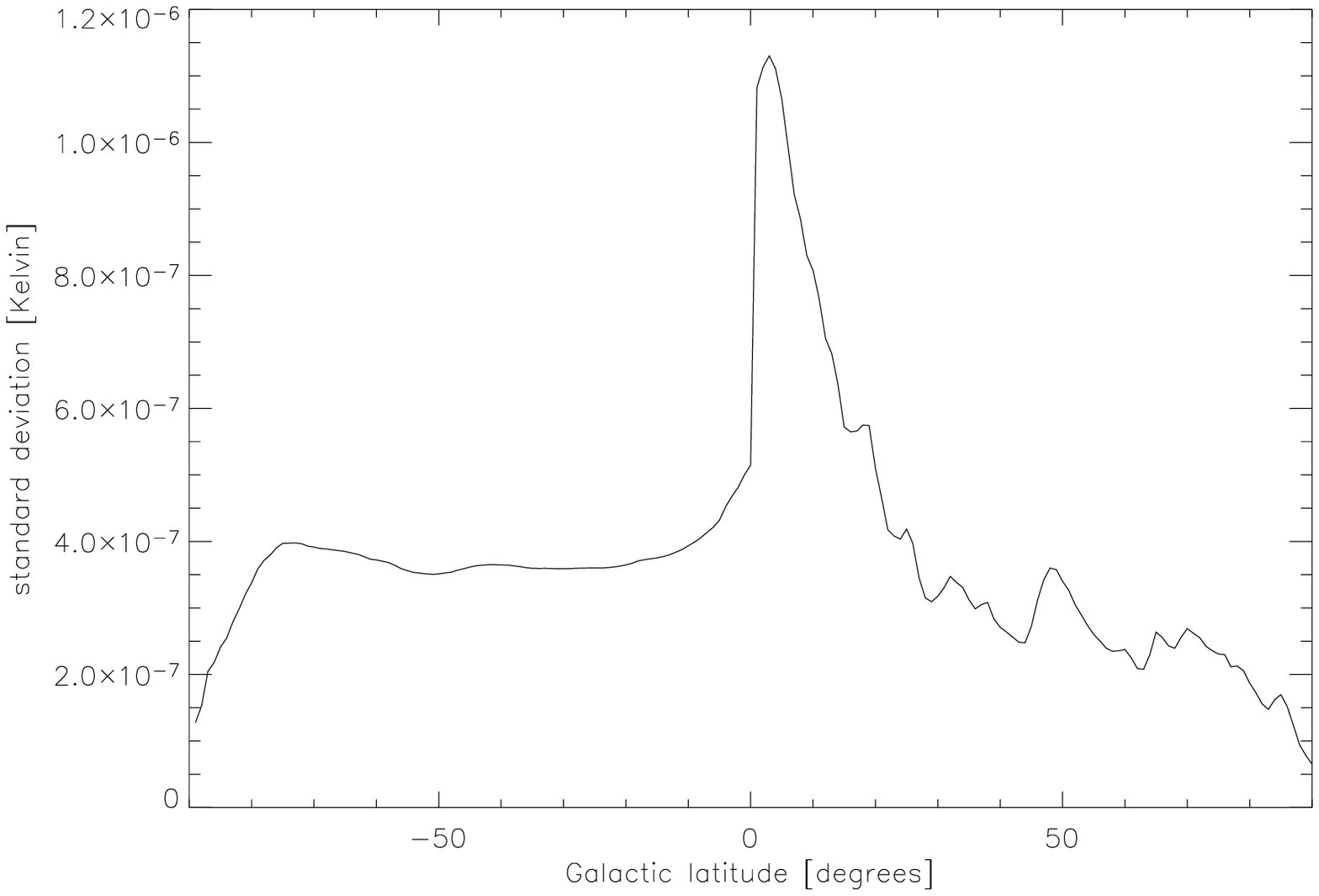}
\includegraphics[width=6cm,height=4.3cm]{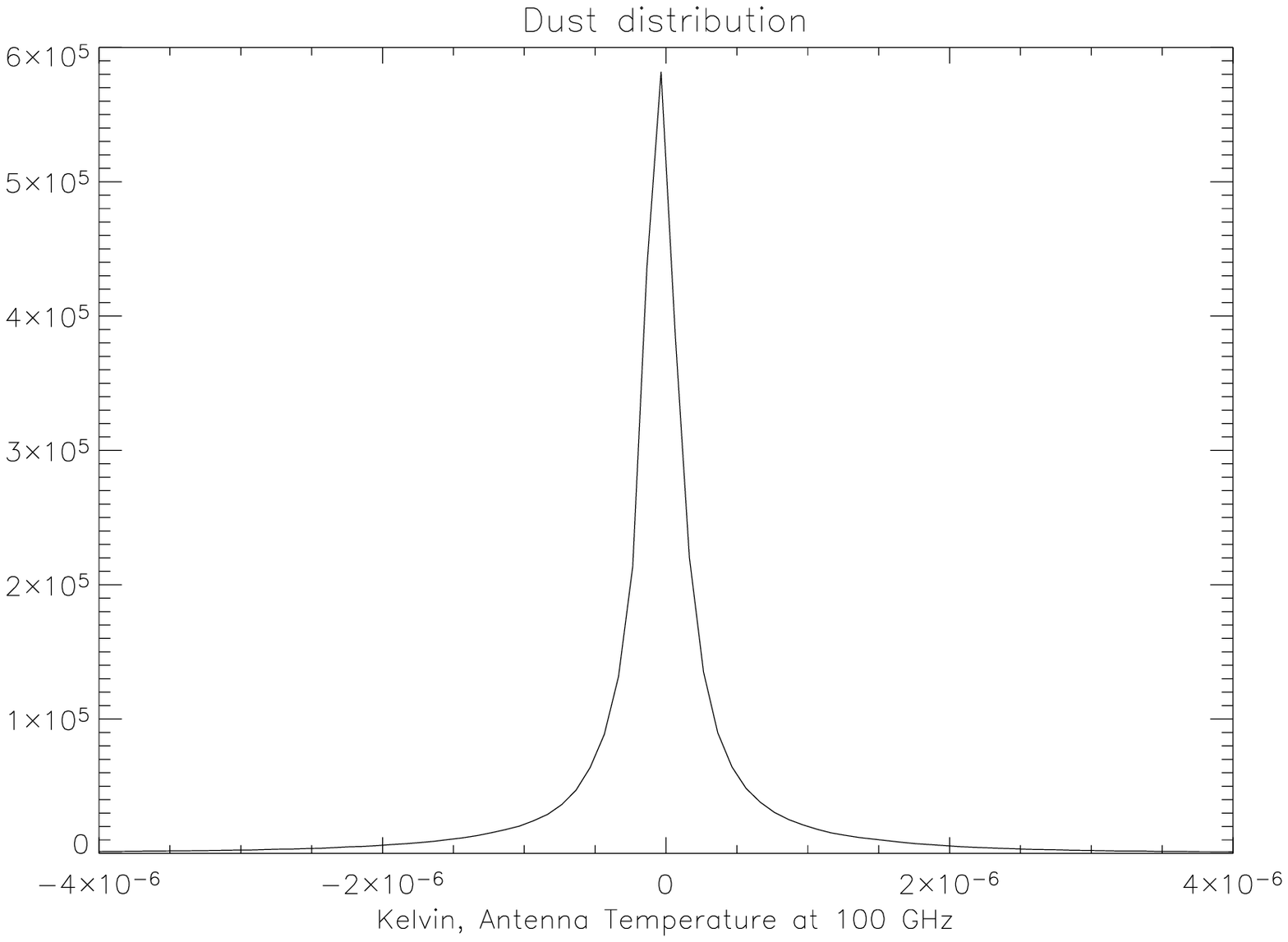}
\includegraphics[width=6cm,height=4.3cm]{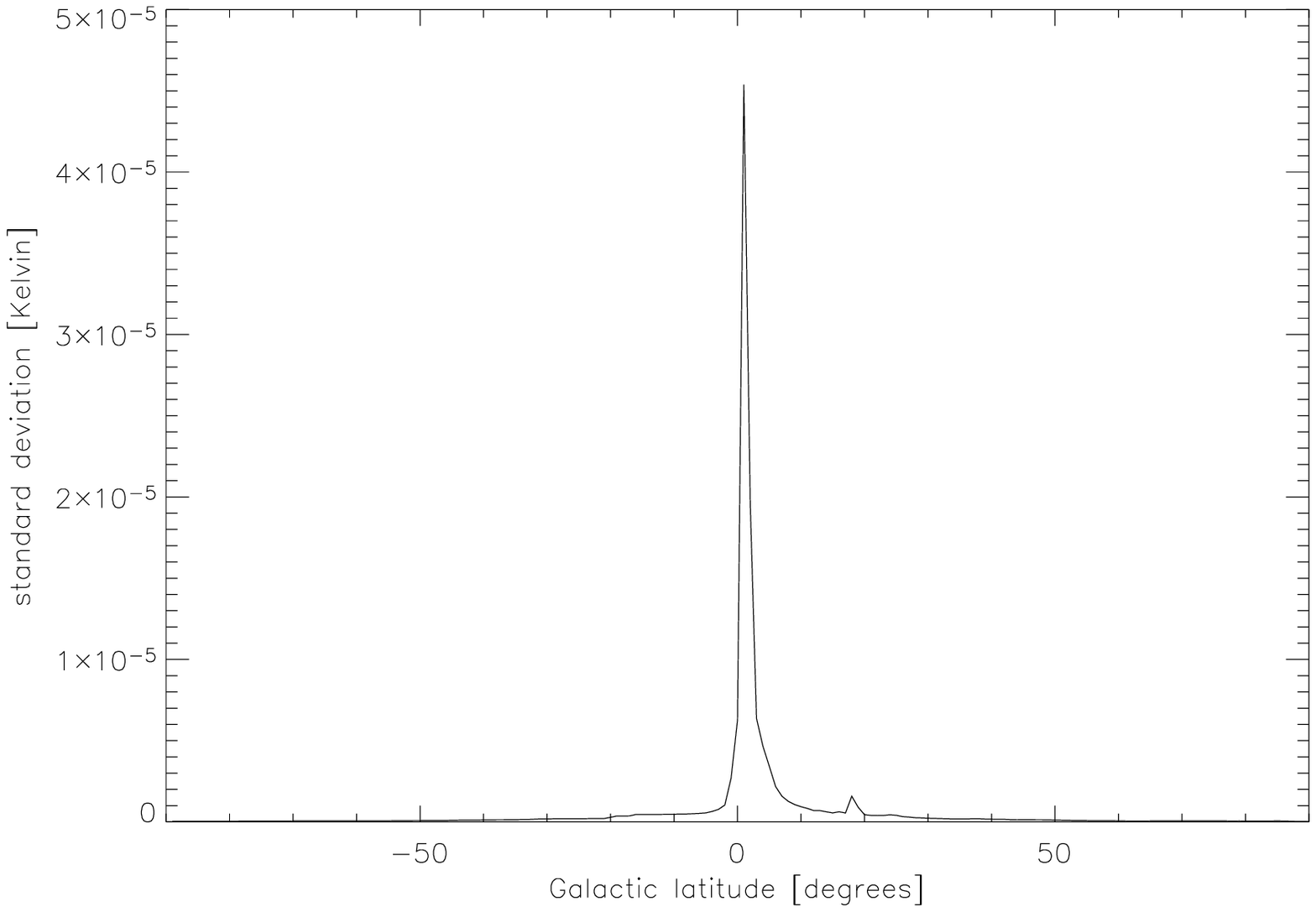}
\caption{Signal distribution for the Stokes parameter $Q$ (left) and Galactic 
latitude dependence (right) for $S_{G}$ (top), $S_{B}$ (middle) and dust 
(bottom), at 100 GHz.}
\label{histo}
\end{figure}

The $Q$ Stokes parameter template\footnote{Unless otherwise specified, 
the maps and power spectra we show are in Kelvin, antenna temperature.} 
obtained by \citet{GIARD}, hereafter 
$S_{G}$, is shown in figure \ref{syndustsky}, together with the spectral index 
template, upper and lower left panels, respectively; we also show the 
$S_{B}$ template obtained rescaling $S_{G}$ to match the power spectrum 
predicted by \citet{BACCI2}, upper right panel; we also show the $Q$ template 
for the polarised thermal dust emission as described above, in the right 
lower panel. $S_{G}$, $S_{B}$ and dust are shown in a non-linear scale in 
order to highlight the behavior at high Galactic latitudes; the corresponding 
template for $U$ is qualitatively equivalent; they have been 
extrapolated to 100 GHz, in antenna temperature, using the spectral 
index template for synchrotron and the latest recipe by \citet{FINKB} for 
dust. By visual inspection, the $S_{B}$ model appear to be less structured 
on small angular scales with respect to the $S_{G}$ and the dust template; 
the latter has the sharpest decrease with Galactic latitude, as we discuss 
below. 

Before quantifying the contamination to the CMB polarisation 
angular power spectrum, it is useful to give a look at the whole 
foreground statistics and sky distribution, plotted in figure 
\ref{histo}; from top to bottom, the left and right panels represent 
signal distribution and root mean square as a function of $b$ 
for $S_{G}$, $S_{B}$ and dust. The signal distributions are 
far from Gaussianity, and exhibit a marked super-Gaussian behavior, 
as the total intensity maps from which all these templates were 
derived. Also, as we anticipated, the dust and $S_{B}$ templates 
present the strongest and weakest dependence on the Galactic latitude, 
respectively. For the dust, the reason is the assumed perfect 
correlation between total and polarised intensity. For $S_{B}$, 
as we see now more in detail, most of the power lies on the small 
angular scales, which propagate well outside the Galactic plane. 

\begin{figure}
\centering
\includegraphics[width=3.9cm,height=6cm]{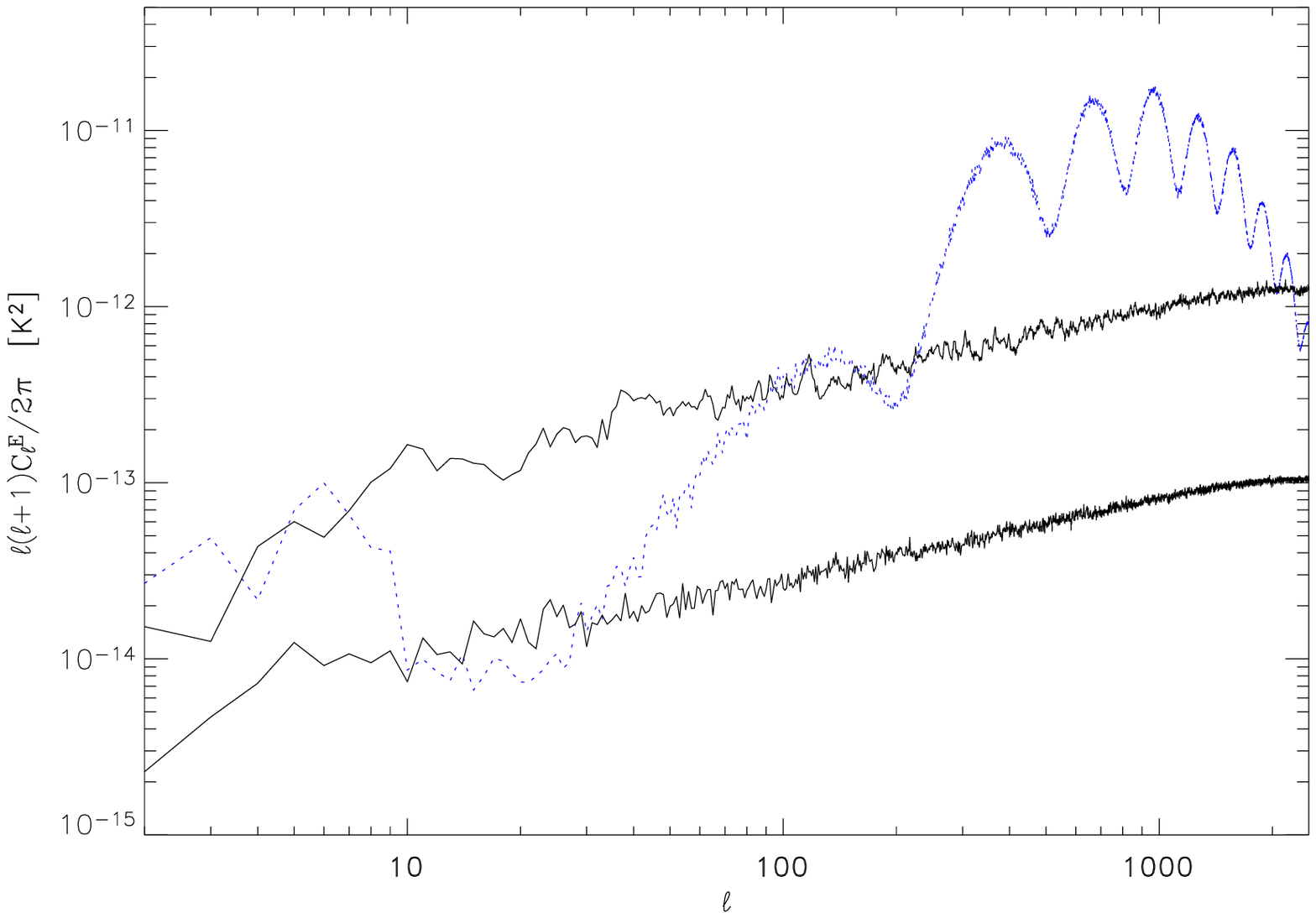}
\includegraphics[width=3.9cm,height=6cm]{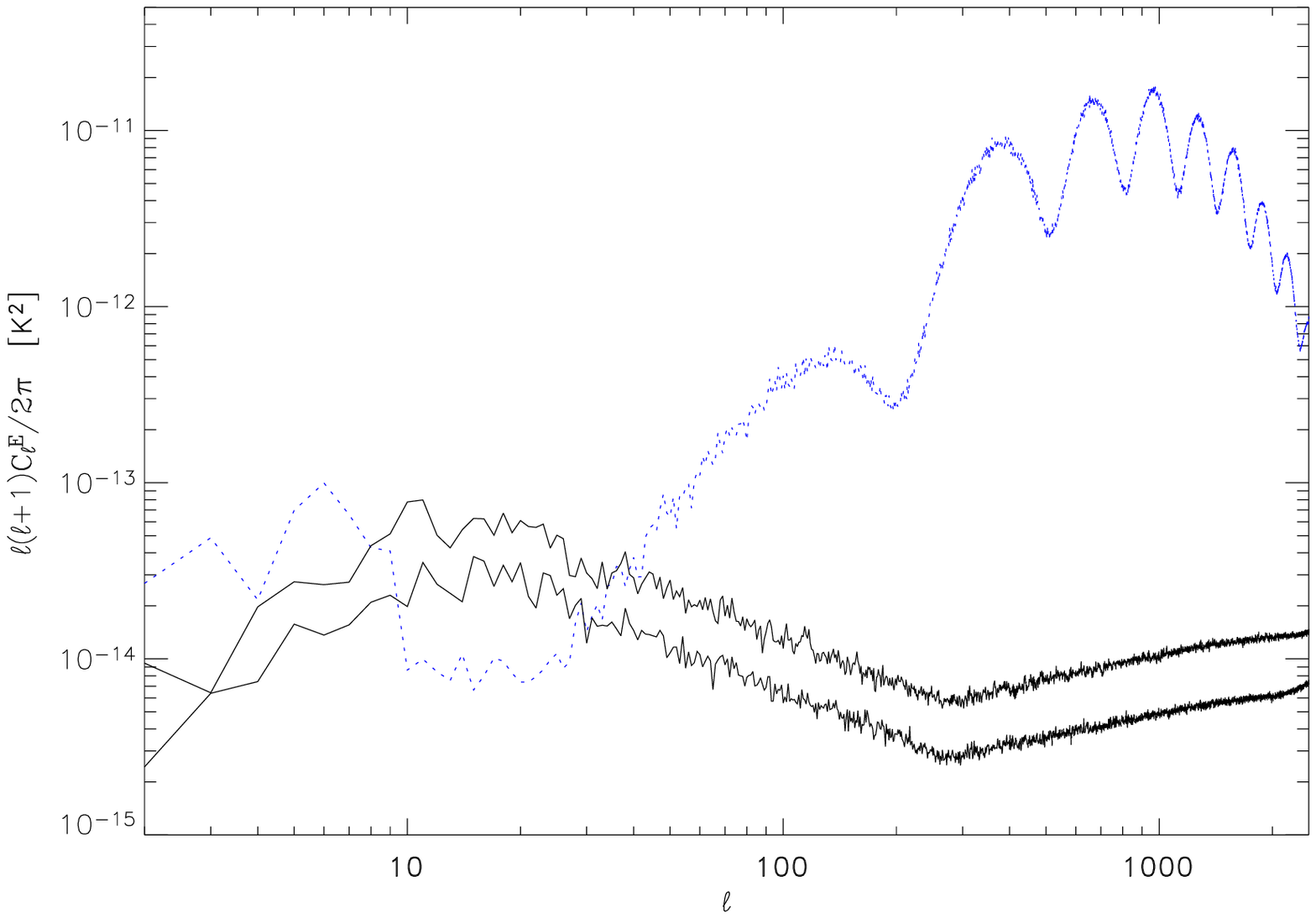}
\includegraphics[width=3.9cm,height=6cm]{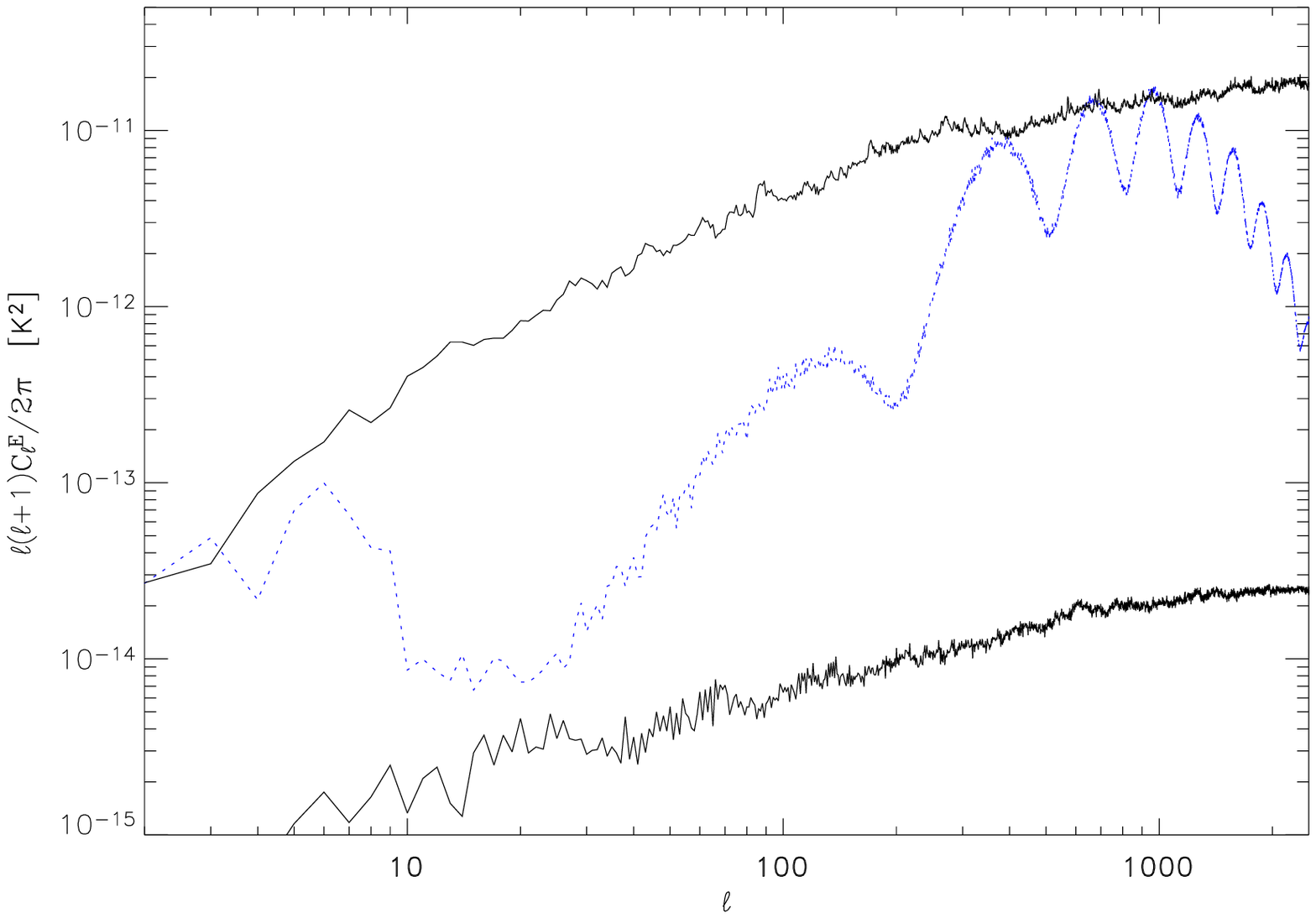}
\includegraphics[width=3.9cm,height=6cm]{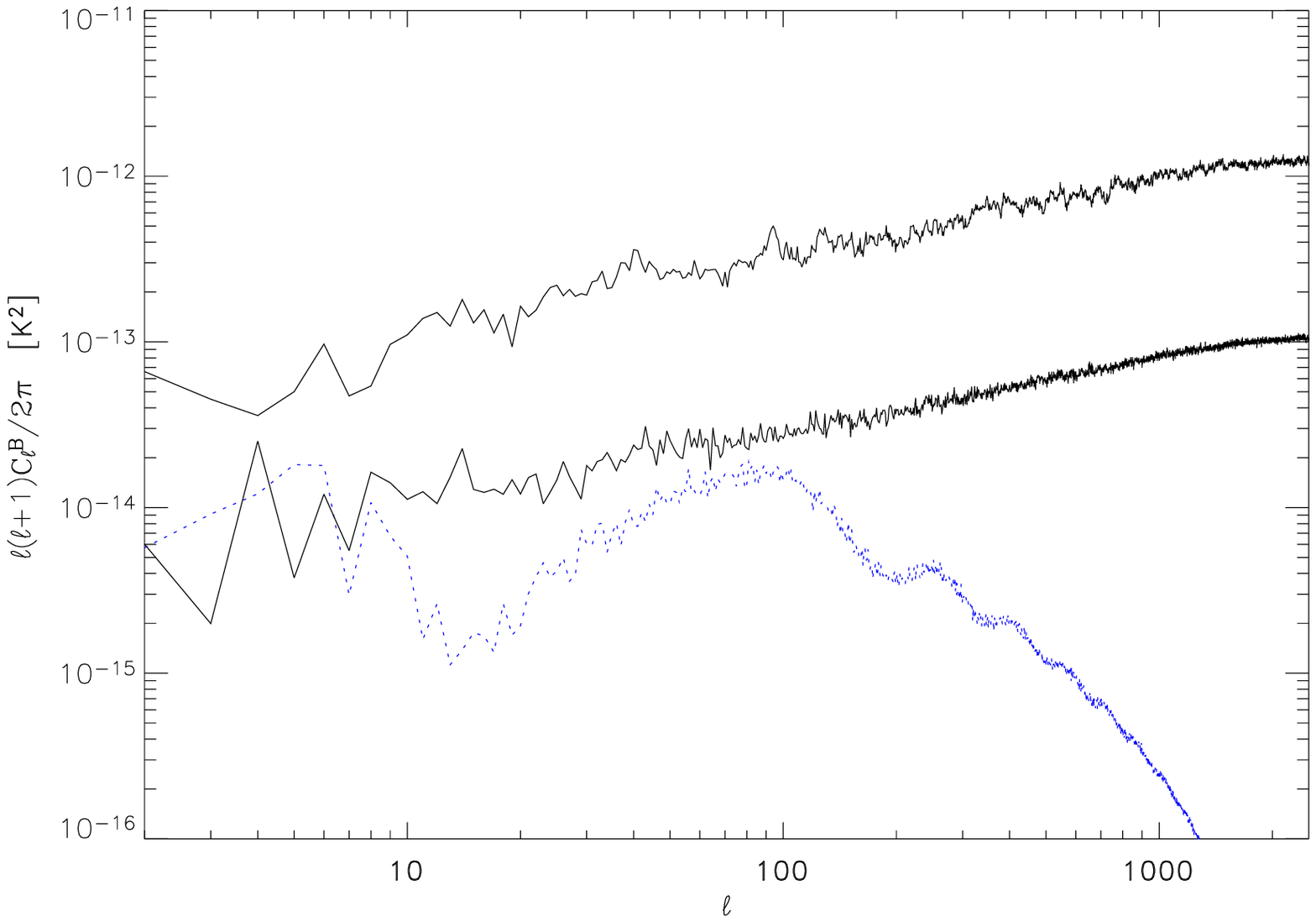}
\includegraphics[width=3.9cm,height=6cm]{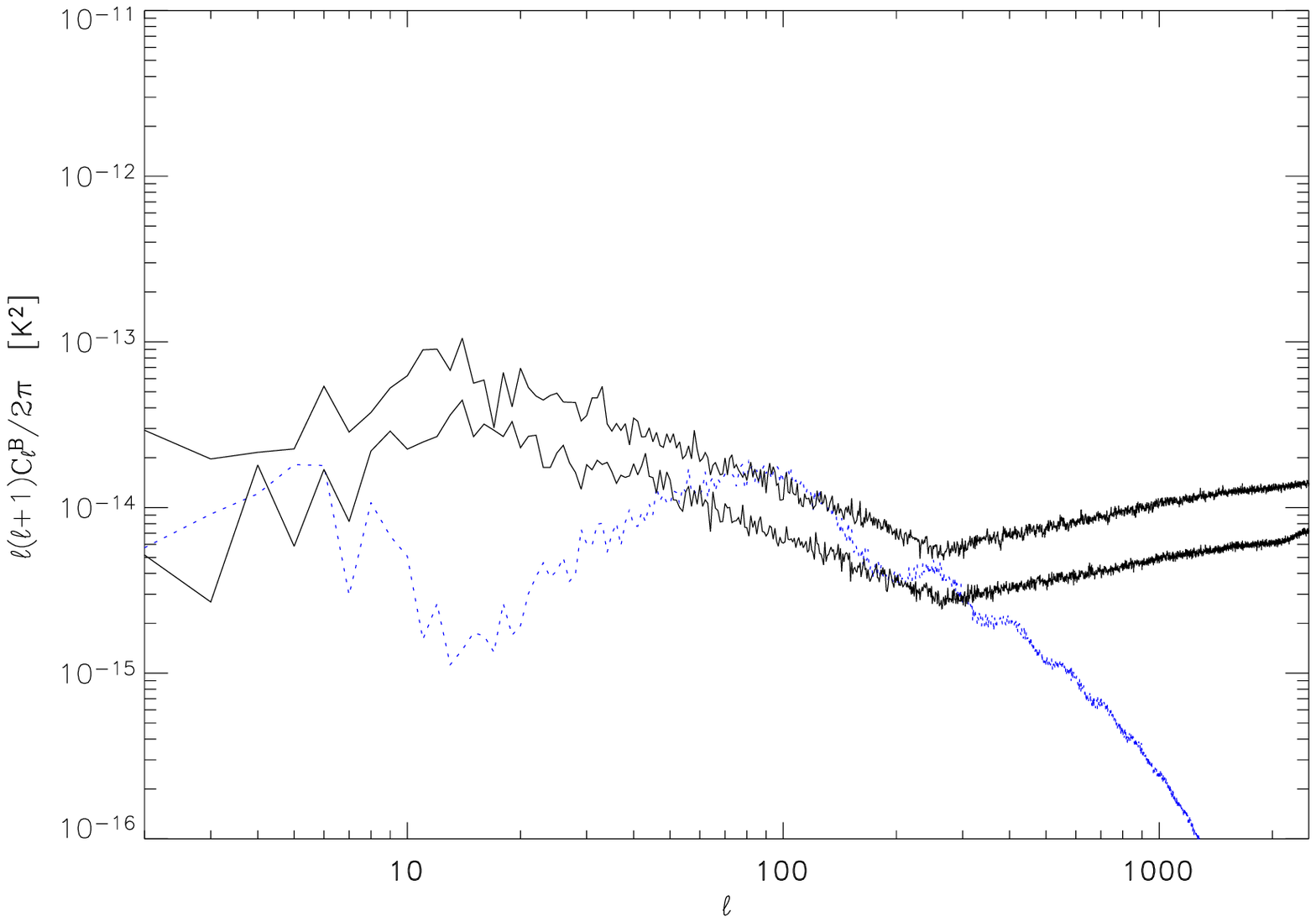}
\includegraphics[width=3.9cm,height=6cm]{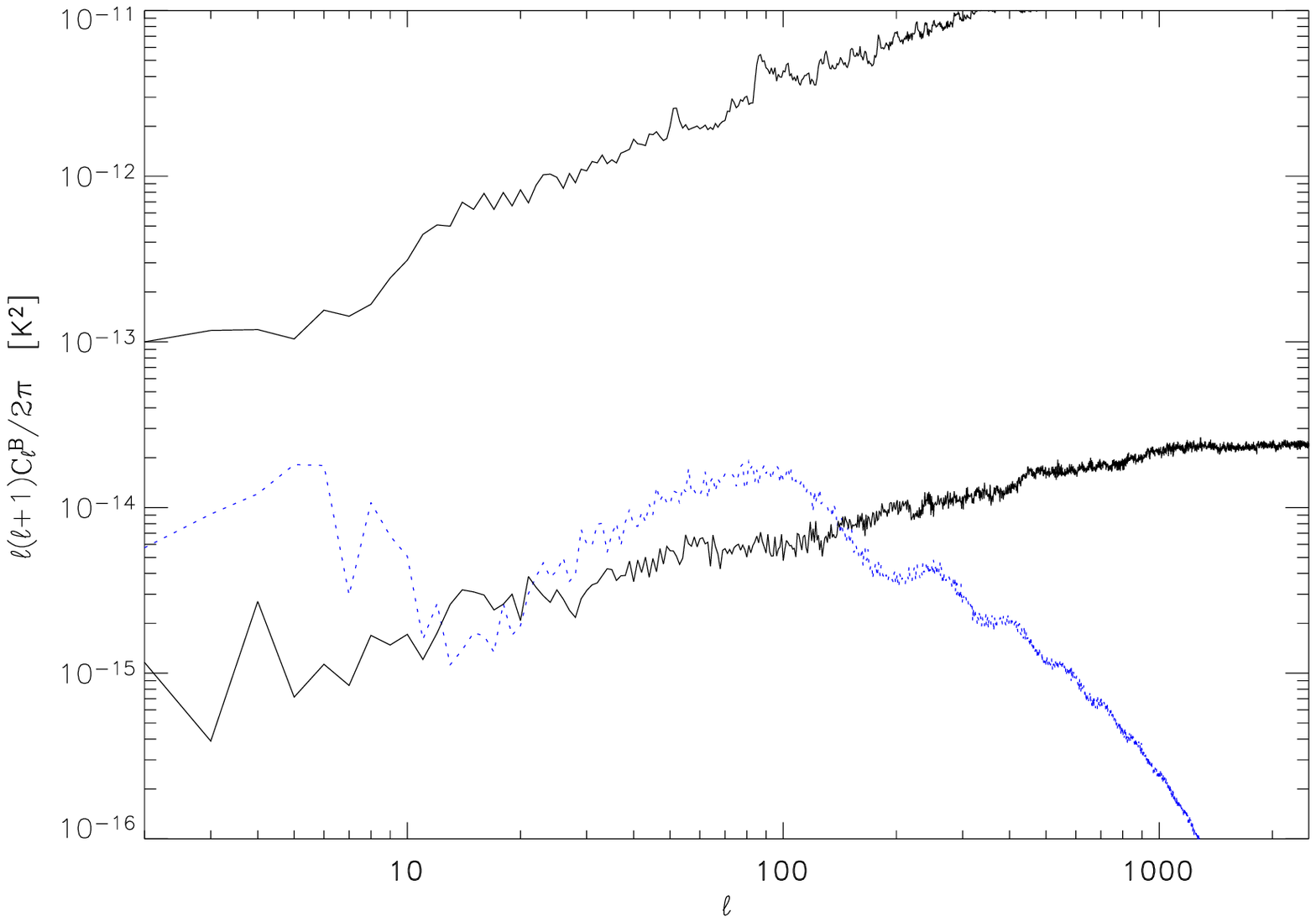}
\includegraphics[width=3.9cm,height=6cm]{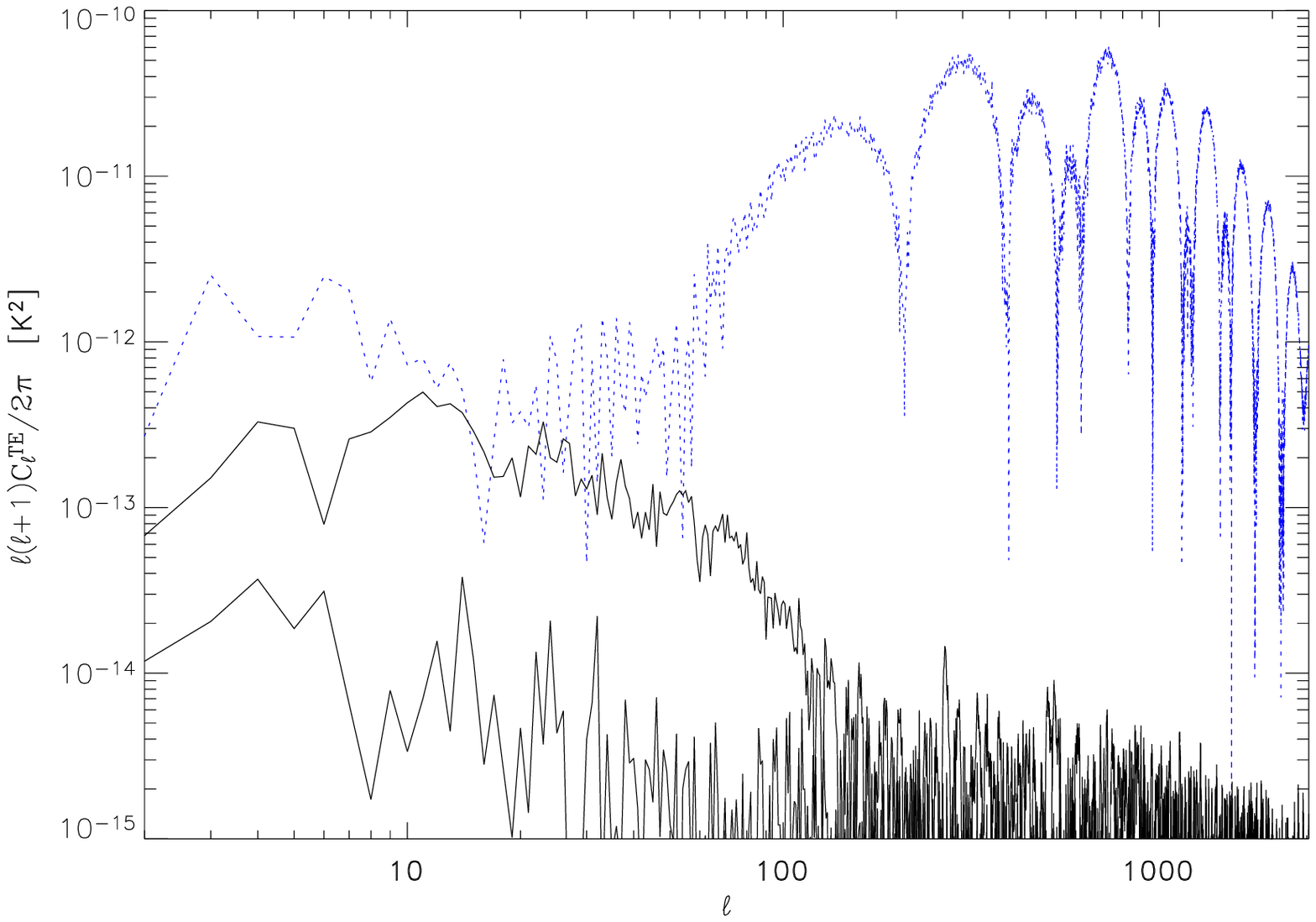}
\includegraphics[width=3.9cm,height=6cm]{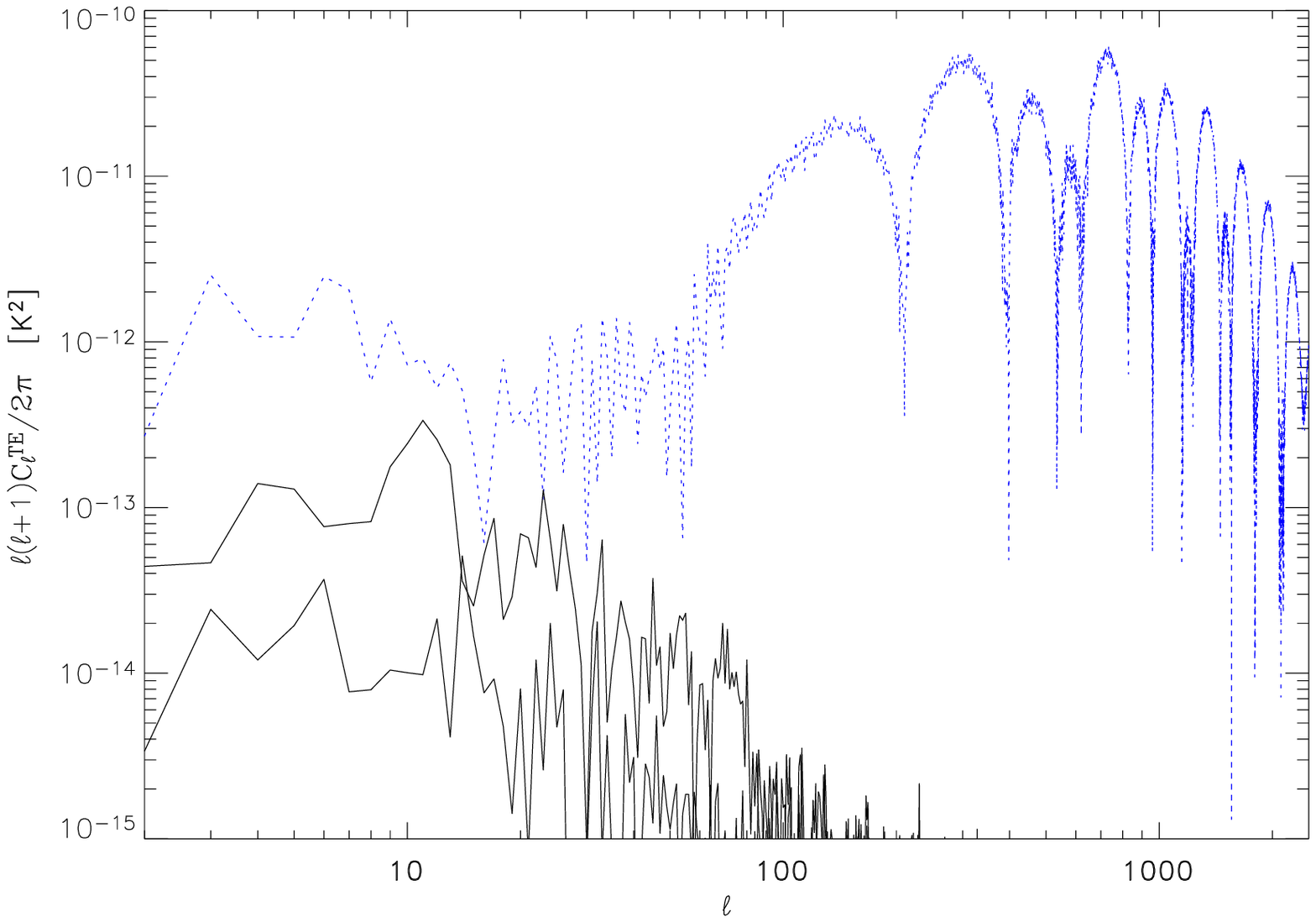}
\includegraphics[width=3.9cm,height=6cm]{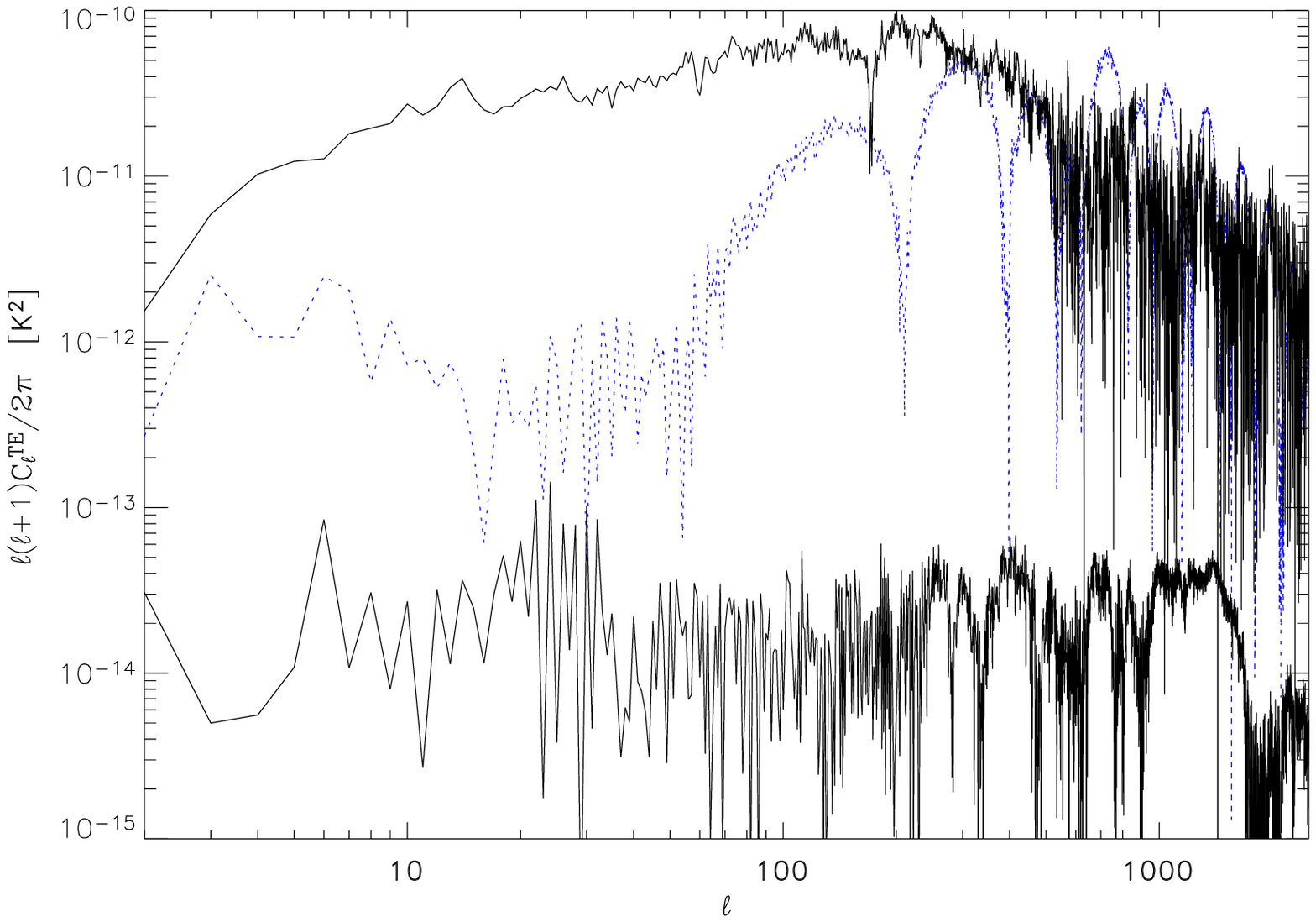}
\caption{Expected angular power spectrum from $S_{G}$ (left panels), 
$S_{B}$ (middle panels) and dust (right panels) with respect to the CMB 
(dotted); in all panels, foreground contamination is calculated on all 
sky and with a Galactic cut for $|b|\le 20^{\circ}$.}
\label{clCMBsyn}
\end{figure}

In Fig. \ref{clCMBsyn} we report the power spectra of the $S_{G}$, 
$S_{B}$ and dust templates, compared to the CMB\footnote{We realize a CMB 
template according to the WMAP data: $h=0.72\ ,\Omega_{\Lambda}=0.7\ ,
\Omega_{b}h^{2}=0.0228\ ,\tau =0.117\ ,n_{S}=0.96$, 
also assuming a tensor to scalar ratio $r=30\%$, rather close to the 
current upper limit \citep[see][and references therein]{BENNE}.}, on all sky 
and with a Galactic cut excluding the region with latitude $|b|\le 20^{\circ}$. 
The plots are at 100 GHz, in antenna temperature. The synchrotron 
frequency scaling according to the template in figure \ref{syndustsky} 
is approximately $[100/\nu\ ({\rm GHz})]^{5.5}$, making the contamination 
rapidly worse at lower frequencies. For the dust, the scaling is 
given by \citet{FINKB}, which in the frequencies of interest here, 
say the interval covered by WMAP and Planck, is approximately 
$\propto\nu^{4.7}/[\exp{(h\nu /k_{B}T_{D})}-1]$, where $h$ and $k_{B}$ 
are the Planck and Boltzmann constant, and the thermodynamical dust 
temperature $T_{D}\simeq 18$ Kelvin; the departure from a pure 
black body behavior is due to the heating that the interstellar 
medium gets from stars \citep[see][and references therein]{PRUNE}. 
The CMB fluctuations in antenna temperature scale as the 
derivative of the black body brightness  
$\propto\nu^{2}\exp{(h\nu /k_{B}T_{CMB})}/[\exp{(h\nu /k_{B}T_{CMB})}-1]^{2}$. 
As a general feature, it can be noted how the Galactic emission has almost 
equal power on $E$ and $B$ modes, according to the most natural expectation 
for a non-cosmological signal (Zaldarriaga 2001). \\
On a full sky analysis, the polarised dust dominates. 
This is likely to be the case if the observed $3-5\%$ polarisation degree 
really holds for all scales and the polarisation angle statistics is not too 
far from the synchrotron one. On the other hand, this signal does show the 
most abrupt drop when the Galaxy is cut out, which cleans almost completely 
the CMB $E$ and $TE$ modes, remaining however substantial for the $B$ ones; 
note that this case agrees remarkably well with the level quoted by 
\citet{PRUNE} at intermediate Galactic latitudes. \\ 
On all sky, the synchrotron contamination is severe on large angular scales, 
almost covering the CMB reionization bump for $E$ and $B$, being 
somewhat less important for $TE$. On $E$, the $S_{B}$ signal drops rapidly 
below the CMB at a few tens in $\ell$, while $S_{G}$ affects the first 
CMB $E$ acoustic oscillation. An important point is that the Galactic 
cut is not as effective as for the dust: the reason is the weaker Galactic 
latitude dependence for synchrotron, as it is evident in figure \ref{histo}. 
As a result, the CMB $B$ modes are likely to be severely contaminated even 
when the Galaxy is cut out. 

We conclude that our present knowledge of the diffuse Galactic polarisation 
emission indicate that CMB $TE$ and $E$ modes are relatively clean, at least 
at medium and high Galactic latitudes and at frequencies around 100 GHz, 
while the $B$ modes are likely to be severely contaminated, at all 
frequencies and Galactic latitudes. 

\section{Component Separation}
\label{compsep}

Despite of the severe foreground contamination, the CMB $B$ modes 
are expected to carry most important details on the nature of 
cosmological perturbations, as we stressed in the Introduction. 
For this reason it is worth to study data analysis techniques capable 
to clean the CMB from foreground contamination. This is feasible by 
exploiting the statistical differences of CMB and diffuse foregrounds, 
starting from a wide enough multi-frequency coverage of a given 
observation. The data can be sketched as 
\begin{equation}
\label{data}
{\bf x}(n_{f},n_{p})={\bf A}(n_{f},n_{c})\cdot{\bf s}(n_{c},n_{p})+
{\bf n}(n_{f},n_{p})\ ,
\end{equation}
where ${\bf x}$ represents the data at the $n_{f}$ frequencies, ${\bf s}$ 
the $n_{c}$ components corresponding to each emission, and ${\bf n}$ 
the instrumental noise; ${\bf A}$ is the mixing matrix, scaling and mixing 
the emissions ${\bf s}$ at any observed frequency, while $n_{p}$ is the 
number of pixels of the observation, or equivalently the number of harmonic 
coefficients if the analysis is performed in the spectral domain. 
Note that in general the angular resolution varies among the frequency 
channels. The target is to recover ${\bf s}$ from ${\bf x}$; as we saw 
in the previous Section, the problem is undetermined since we know 
{\bf A} only partially, and the guesses on {\bf s} are not reliable 
for stabilizing the inversion of the relation (\ref{data}), either if 
the noise statistical properties are very well known, or even in a 
noiseless case. 
On the other hand, a well established expectation is that the 
CMB and foreground emission are statistically independent. It has been 
recently shown that this feature can be exploited to design ``blind" 
component separation tools, i.e. able to recover ${\bf s}$ even without 
any prior on {\bf s} or {\bf A}; a first algorithm of this class 
was proposed for use in astrophysical component separation by \citet{BACCI1}
based on the novel concept in signal processing science, the Independent 
Component Analysis: that works by finding a ${\bf W}(n_{c},n_{f})$ matrix 
which applied to ${\bf x}(n_{f},n_{c})$ in (\ref{data}) brings 
to ${\bf s}+{\bf W}{\bf n}$. The criterion to find ${\bf W}$ is to look 
for the statistically independent patterns in ${\bf x}$: that is clearly 
not unique and can be specialized for the problem at hand. A fast version 
({\ica}) adapted for satellite observations has been proposed by 
\citet{MAINO1}; this technique has been further developed to account for 
any available and reliable prior, and successfully applied to the real COBE 
data by \citet{MAINO2}. Recently, a different and promising blind technique 
based on a spectral matching of real and modeled data \citet{DELAB}. 

\begin{figure}
\centering
\includegraphics[width=6cm,height=6.4cm]{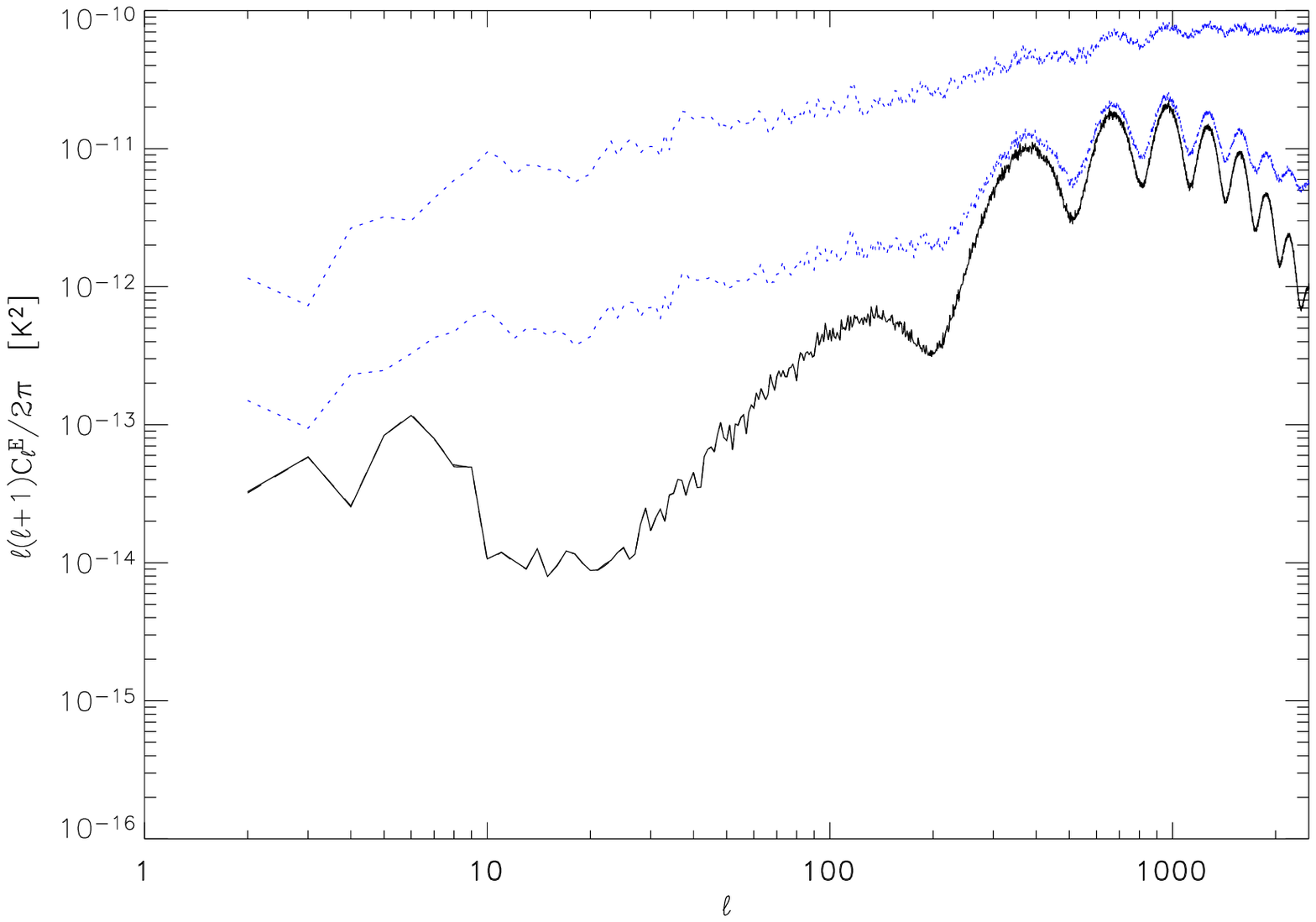}
\includegraphics[width=6cm,height=6.4cm]{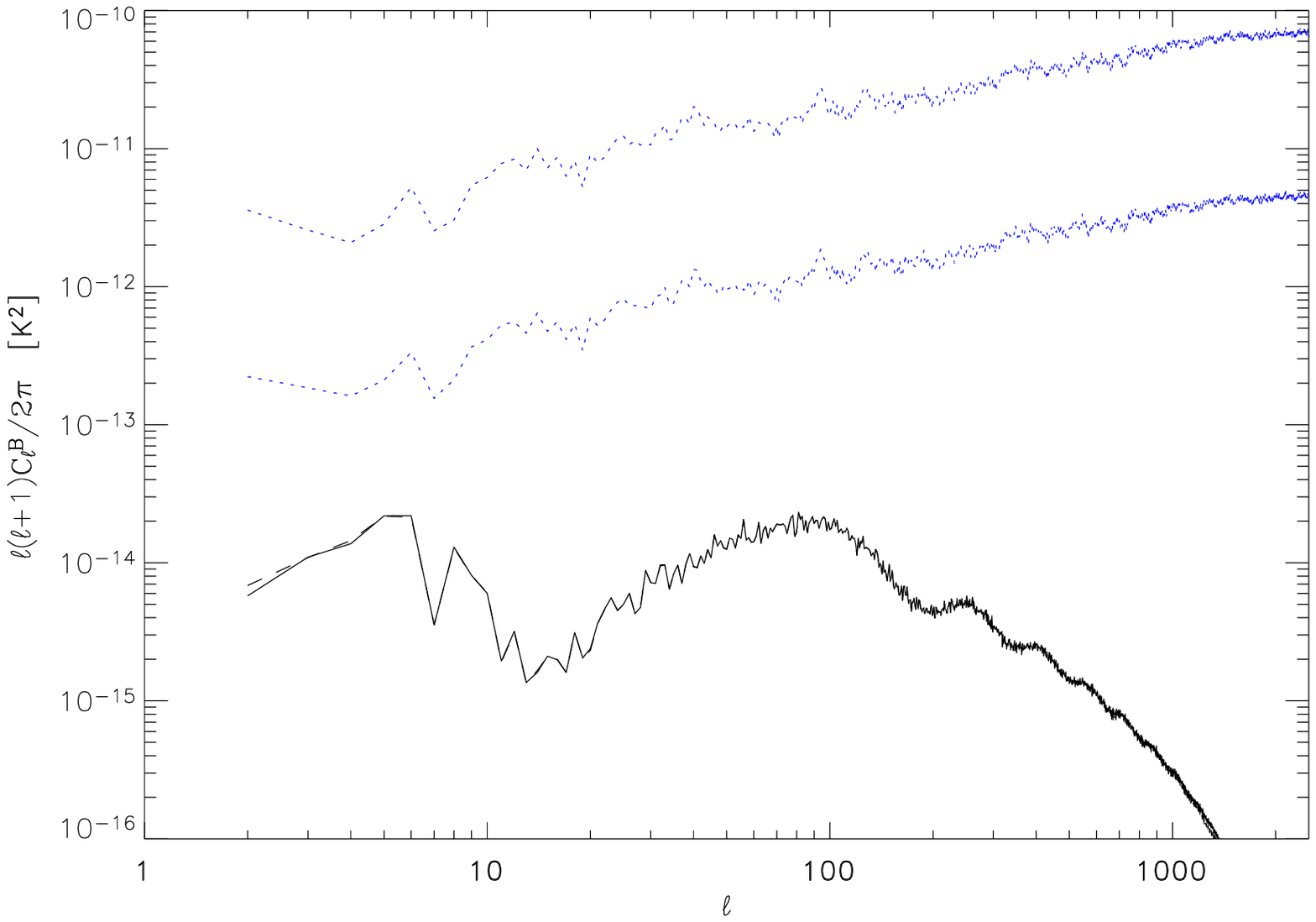}
\caption{$E$ and $B$ modes of the original sky at 50 and 80 GHz 
(respectively high and low dotted curves) and recovered CMB $E$ and $B$ modes 
after component separation (solid), at 80 GHz; the true $CMB$ (dashed), 
almost indistinguishable from the recovered one, is also plotted.}
\label{compsepresults}
\end{figure}

The first application of component separation to polarisation radiation 
has been performed by \citet{BACCI3}, with a {\ica} approach: 
separation is conducted on $Q$ and $U$ Stokes parameters separately or 
considered as a single signal provided that they obey the same statistics; 
we refer to that work for any detail on the algorithm procedure. Here we 
show a very simple application, with rather impressive results, to highlight 
the potentialities of this technique. We consider a noiseless mixture of 
polarised CMB and synchrotron $S_{G}$, scaling uniformly in frequency, 
at 50 and 80 GHz, on all sky and pixel size of about 3.5 arcmin.; note that 
we do not refer to any operating or planned experiment, and choose the 
frequencies among those in which CMB and synchrotron are roughly comparable. 
The two channels are input to the {\ica} to recover synchrotron and CMB 
separately. Figure \ref{compsepresults} show the input sky angular power 
spectra for $E$ (left) and $B$ (right) together with the spectra of the 
recovered CMB as output by {\ica}. As it is evident, a CMB major cleaning 
was achieved: the dotted line is the input sky $B$ power, while the 
recovered CMB (solid) line is actually almost indistinguishable from the 
input one (dashed). We stress that this is without the use of any prior, i.e. 
not knowing any piece of {\bf A} or ${\bf s}$. The algorithm knows only the 
data at the two frequencies, and looks for the most statistically independent 
components in those data. In addition, applying this technique in particular 
is computationally free: the results in figure were obtained in about five 
minutes on an ordinary single processor personal computer. \\
In particular, the CMB $B$ modes have been impressively cleaned: 
interpreting the foreground emission as noise, the signal to noise ratio 
is 0.1 or less, at all multipoles, as it can bee seen in the right panel 
of the figure. CMB $E$ modes have also been cleaned wherever the 
foreground contamination was relevant. The algorithm outputs also 
{\bf A}, which differs from the real one by percent or less. 

The reason of such an excellent performance is just the high level of 
detail of the underlying data. At arcmin. resolution, the maps are almost 
ideal datasets where the algorithm can look for the least statistically 
linked templates present into the maps themselves. Even if all the 
instrumental details were neglected, what we show here should be 
enough to gather attention on these new data analysis techniques, as 
we further stress in the next Section. 

\section{Concluding remarks}
\label{conclud}

In this work we show that at frequencies around 100 GHz the diffuse 
Galactic polarisation emission is likely to be rather weak in comparison to 
the CMB $TE$ and $E$ modes, at least at medium and high Galactic latitudes, 
while the $B$ ones are seriously contaminated on all sky. At 
lower and higher frequencies, the contamination gets rapidly worse because 
of the steep frequency behavior of synchrotron and dust, respectively. 

We stress that the new concepts in signal processing science can be 
exploited in this context, to cancel or greatly reduce such a contamination. 
In particular, the statistical independence of CMB and foreground emission 
is suitable to achieve this goal: we give an impressive example based 
on the {\ica} technique in polarisation \citep{BACCI3}, where CMB $B$ 
modes are recovered out of a foreground emission about ten times stronger, 
without any prior information but their statistical independence, and at an 
almost null computational cost. 

These features represent the ``nominal" capabilities of these algorithms, 
i.e. their performance on ideal datasets: cleaning the cosmological signal 
from substantial foreground contamination, exploiting only their natural 
diversity. This is in our opinion a great achievement, providing the basis 
on which the implementation to any particular observation can be built. 
The latter step, of course, presents challenging issues: the noise level, 
sky distribution and non-stationarity, as well as the non-uniform spectral 
index of the components to recover, are good examples. However, we have 
already the indications of a substantial stability of the nominal 
performance against instrumental features and systematics: \citet{MAINO1} 
and \citet{BACCI3}, respectively for total intensity and polarisation, 
perform a whitening of simulated Planck data in order to get rid of the 
noise bulk in the {\ica} separation process. Most importantly, the results 
of \citet{MAINO2} show that these techniques are flexible enough to be 
applied to real data, the COBE ones, getting outstanding results, namely 
the correct large scale CMB power spectrum amplitude and shape. 

While waiting for the era of large scale CMB polarisation measurements, 
the next obvious target is a successful application to the WMAP data, 
meaning the recovery of CMB data on sky regions affected by Galactic 
contamination which are normally cut out. 

\vskip.1in
The author warmly thanks Gianfranco De Zotti, Giovanna Giardino, 
Eric Hivon, Francesca Perrotta, George Smoot and Radek Stompor for useful 
discussions. The HEALPix pixelisation scheme, available at 
{\tt www.eso.org/healpix}, 
by A.J. Banday, M. Bartelmann, K.M. Gorski, F.K. Hansen, E.F. Hivon, and 
B.D. Wandelt, has been extensively used.


\begin{thebibliography}{}

\bibitem[Baccigalupi\ et\ al.(2000)]{BACCI1}
Baccigalupi C. et al. 2000, MNRAS 318, 769
\bibitem[Baccigalupi\ et\ al.(2001)]{BACCI2}
Baccigalupi C. et al. 2001, A\& A 372, 8
\bibitem[Baccigalupi\ et\ al.(2003)]{BACCI3}
Baccigalupi C. et al. 2003, submitted to MNRAS, 
preprint {\tt astro-ph/0209591}
\bibitem[Bennett\ et\ al.(2003)]{BENNE}
Bennett et al. 2003, ApJ submitted, preprint 
available at {\tt map.gsfc.nasa.gov}
\bibitem[Brouw\ \&\ Spoelstra(1976)]{BROUW}
Brouw W.N., Spoelstra T.A.T., 1976, A\& AS 26, 129
\bibitem[de\ Oliveira-Costa\ et\ al.(2003)]{DEOLI}
de Oliveira-Costa et al. 2003, to be published in the proceedings of 
``The Cosmic Microwave Background and its Polarization", New 
Astronomy Reviews, (eds. S. Hanany and K.A. Olive).
\bibitem[De\ Zotti(2002)]{DEZOT}
De Zotti G., 2002 in Astrophysical Polarized Backgrouds, AIP 
conference proc. 609, S. Cecchini, S. Cortiglioni, R. Sault, and 
C. Sbarra eds., p. 295
\bibitem[Delabrouille\ et\ al.(2003)]{DELAB}
Delabrouille J., Cardoso J.F., Patanchon G. 2003, submitted to MNRAS, 
preprint {\tt astro-ph/0211504}
\bibitem[Duncan\ et\ al.(1999)]{DUNCA}
Duncan A.R., Reich P., Reich W., F$\ddot{\rm u}$rst E., 1999, A\& A 350, 447
\bibitem[Finkbeiner\ et\ al.(1999)]{FINKB}
Finkbeiner D.P., Davis M., Schlegel D.J. 1999, ApJ 524, 867
\bibitem[Giardino\ et\ al.(2002)]{GIARD}
Giardino G. et al. 2002, A\& A 387, 82
\bibitem[Haslam\ et\ al.(1982)]{HASLA}
Haslam C.G.T. et al. 1982, A\&A S 47, 1
\bibitem[Hu\ et\ al.(1998)]{HU}
Hu W., Seljak U., White M., Zaldarriaga M. 1998, Phys.Rev.D 57, 3290
\bibitem[Kovac\ et\ al.(2002)]{KOVAC}
Kovac J. et al. 2002, ApJ submitted, {\tt astro-ph/0209478}
\bibitem[Lazarian\ \&\ Prunet(2002)]{LAZAR}
Lazarian A., Prunet S. 2002, in Astrophysical Polarized Backgrouds, 
AIP conference proc. 609, 
S. Cecchini, S. Cortiglioni, R. Sault, and C. Sbarra eds., p. 32
\bibitem[Maino\ et\ al.(2002)]{MAINO1}
Maino D. et al. 2002, MNRAS 334, 53
\bibitem[Maino\ et\ al.(2003)]{MAINO2}
Maino D. et al. 2003, submitted to MNRAS, preprint 
{\tt astro-ph/0303657}
\bibitem[Ponthieu\ et\ al.(2003)]{PONTH}
Ponthieu et al. 2003, to be published in the proceedings of 
``The Cosmic Microwave Background and its Polarization", New 
Astronomy Reviews, (eds. S. Hanany and K.A. Olive).
\bibitem[Prunet\ et\ al.(2000)]{PRUNE}
Prunet S., Sethi S.K., Bouchet F.R., 2000, MNRAS 314, 348
\bibitem[Tegmark\ et\ al.(2000)]{TEGMA}
Tegmark M., Eisenstein D.J., Hu W., de Oliveira-Costa A., 2000, 
ApJ 530, 133
\bibitem[Tucci\ et\ al.(2000)]{TUCCI1}
Tucci M. et al. 2000, New Astronomy 5, 181
\bibitem[Tucci\ et\ al.(2002)]{TUCCI2}
Tucci M. et al. 2002, ApJ 579, 607
\bibitem[Uyaniker\ et\ al.(1999)]{UYANI}
Uyaniker B. et al. 1999, A\& AS 138, 31
\bibitem[White(2003)]{WHITE}
White M., 2003, to be published in the proceedings of 
``The Cosmic Microwave Background and its Polarization", New 
Astronomy Reviews, (eds. S. Hanany and K.A. Olive).
\bibitem[Zaldarriaga(2001)]{ZALDA}
Zaldarriaga M., 2001, Phys.Rev.D 64, 103001

\end{thebibliography}
\end{document}